\documentclass[amsmath,amssymb,prl,preprint,superscriptaddress]{revtex4-1}
\usepackage{graphicx}
\usepackage{dcolumn}
\usepackage{bm}
\usepackage{natbib}
\usepackage{lineno}
\usepackage{color} 
\definecolor{dickred}{rgb}{0,0,0}

\setcitestyle{super}

\begin{document}

\title{Local noise in a diffusive conductor}
\author{E.S.~Tikhonov}
\affiliation{Moscow Institute of Physics and Technology, Dolgoprudny, 141700 Russian Federation}
\affiliation{Institute of Solid State Physics, Russian Academy of
Sciences, 142432 Chernogolovka, Russian Federation}
\author{D.V.~Shovkun}
\affiliation{Moscow Institute of Physics and Technology, Dolgoprudny, 141700 Russian Federation}
\affiliation{Institute of Solid State Physics, Russian Academy of
Sciences, 142432 Chernogolovka, Russian Federation}
\author{D.~Ercolani}
\affiliation{NEST, Istituto Nanoscienze -- CNR and Scuola Normale Superiore, Piazza S. Silvestro 12, I-56127 Pisa, Italy}
\author{F.~Rossella}
\affiliation{NEST, Istituto Nanoscienze -- CNR and Scuola Normale Superiore, Piazza S. Silvestro 12, I-56127 Pisa, Italy}
\author{M.~Rocci}
\affiliation{NEST, Istituto Nanoscienze -- CNR and Scuola Normale Superiore, Piazza S. Silvestro 12, I-56127 Pisa, Italy}
\author{L.~Sorba}
\affiliation{NEST, Istituto Nanoscienze -- CNR and Scuola Normale Superiore, Piazza S. Silvestro 12, I-56127 Pisa, Italy}
\author{S.~Roddaro}
\affiliation{NEST, Istituto Nanoscienze -- CNR and Scuola Normale Superiore, Piazza S. Silvestro 12, I-56127 Pisa, Italy}
\author{V.S.~Khrapai} 
\email{Correspondence to dick@issp.ac.ru}
\affiliation{Moscow Institute of Physics and Technology, Dolgoprudny, 141700 Russian Federation}
\affiliation{Institute of Solid State Physics, Russian Academy of
Sciences, 142432 Chernogolovka, Russian Federation}

\begin{abstract}
\textbf{The control and measurement of local non-equilibrium configurations is of utmost importance in applications on energy harvesting, thermoelectrics and heat management in nano-electronics. This challenging task can be achieved with the help of various local probes,
prominent examples including superconducting  \cite{Pothier1997,Giazotto_Pekola_2006} or quantum dot based \cite{Altimiras2009,Hoffmann2009,Venkatachalam2012} tunnel junctions, classical \cite{Wu2013} and quantum \cite{Prokudina2014} resistors, and Raman thermography \cite{Doerk2010,Yazji2015}. Beyond time-averaged properties, valuable information can also be gained from spontaneous fluctuations of current (noise). 
From these perspective, however, a fundamental constraint is set by current conservation, which makes noise a characteristic of the whole conductor, rather than some part of it. Here we demonstrate how to remove this obstacle and pick up a local noise temperature of a current biased diffusive conductor with the help of a miniature noise probe. This approach is virtually noninvasive \textcolor{dickred}{for the electronic energy distributions} and extends primary local measurements towards strongly non-equilibrium regimes.}
 
\end{abstract}

\maketitle

Charge transport in electronic conductors is stochastic by origin, as such is the motion of each individual carrier~\cite{Blanter2000}. 
Mean squared current fluctuation (current noise) contains valuable information about quasiparticle charge~\cite{Ronen2016}, transport statistics~\cite{Choi2015,Matsuo2015,Kumada2015} and is particularly useful for Johnson-Nyquist \cite{White1996} or non-equilibrium \cite{Roukes1985,Strunk1998,Spietz2003,PhysRevX.2.031006,Laitinen2014} noise thermometry and even for dynamic measurements of the inelastic relaxation rates~\cite{Reulet_2016arXiv}. Microscopically, the noise originates from the random fluctuations $\delta n_k$ of the occupation numbers of the electronic quantum states. Altogether, such fluctuations determine the local noise temperature $T_N$ of the electronic system and give rise to a spontaneous current fluctuation $\overline{\delta I_s^2}\propto T_N$. This fluctuation represents a random current source acting on the conductor at a given location. \textcolor{dickred}{While in the ballistic regime or in quantum confined conductors $\delta n_k$  is preserved along the device, in general such random sources are mutually uncorrelated at distances exceeding the carrier mean-free path.} Current conservation, however, dictates that the actually measurable fluctuation is the same within any cross-section of the conductor and averaged over all sources~\cite{Nagaev1992}, $\overline{\delta I^2}\propto\int\overline{\delta I_s(x)^2}dx/L$, where $x$ is the coordinate along the conductor and $L$ is its length. This fundamental constraint makes conventional noise measurements sample-averaged and hides the information about the local noise temperature. 

\begin{figure}[h]
\begin{center}
\vspace{10mm}
  \includegraphics[width=0.9\columnwidth]{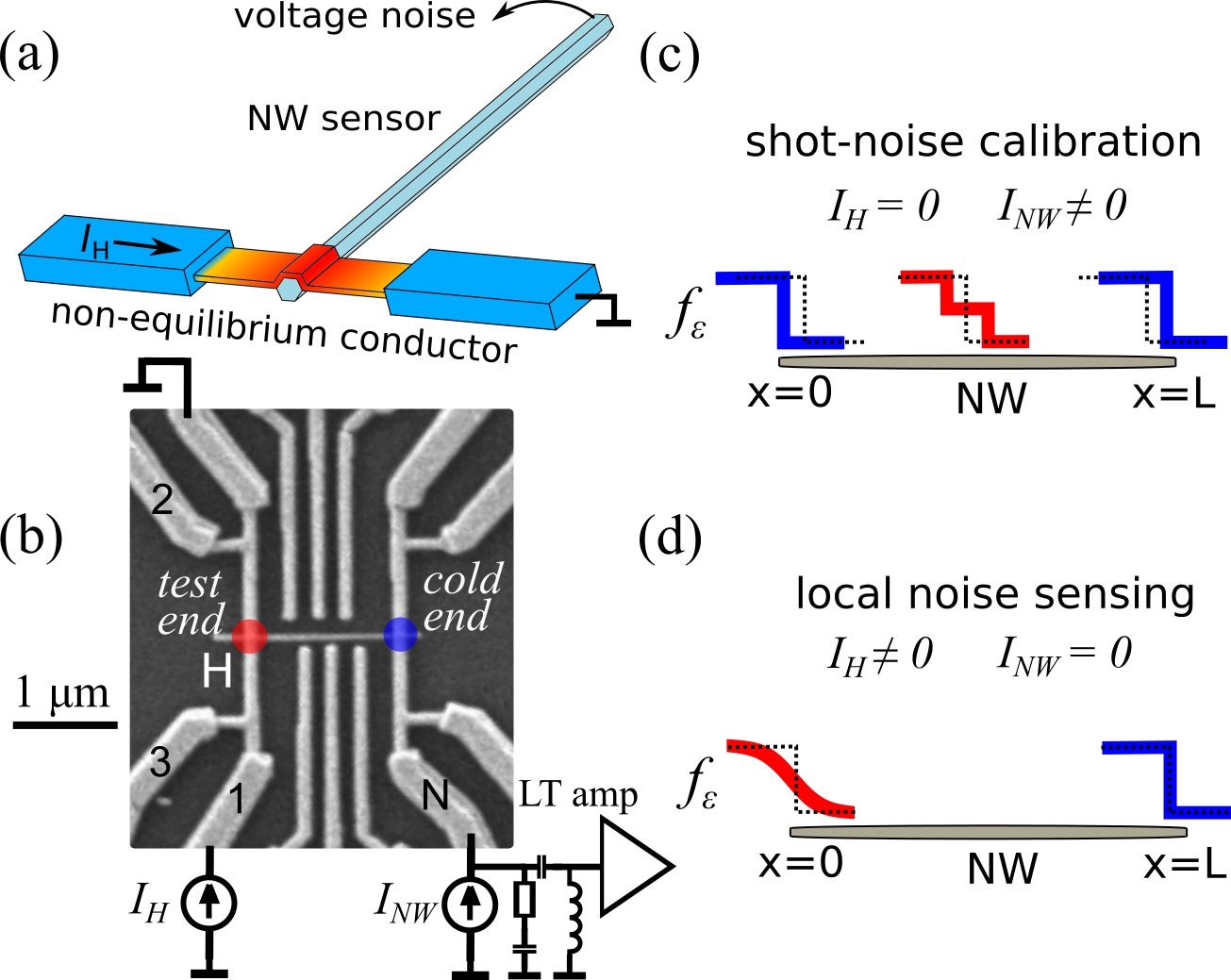}
   \end{center}
  \caption{Concept of the local noise measurement. (a) -- local NW-based noise sensor (thin rod) in contact with the probed non-equilibrium  conductor. Color gradient depicts a spatial inhomogeneity of the electronic energy distribution in the conductor. (b) --
scanning electron micrograph image of one of our nominally identical devices. The NW and surrounding metallic nanostructure are shown in light-gray. The red and blue spots mark, respectively, the test end and the opposite cold end of the NW. Electric layout shows the connections at the input of the low-temperature rf-amplifier used for noise measurements as well as the dc circuit used to drive the constriction H out of equilibrium. (c) and (d) -- electronic energy distributions $f_\varepsilon$ at different positions inside the NW corresponding to the shot noise calibration procedure and the local noise measurement (thick solid lines), respectively. Thinner dotted lines correspond to $f_\varepsilon$ in thermal equilibrium at zero temperature. }\label{sketch}
\end{figure}

To overcome this constraint, we propose a concept of a noise sensor depicted in Fig.~\ref{sketch}a. The sensor, represented by a thin rod, is brought into contact with a current biased conductor, in which a spatially inhomogeneous non-equilibrium state is indicated by a color gradient. The non-equilibrium electronic energy distribution at the contact position is picked up by the sensor and gives rise to measurable excess noise at a zero net current. A proper non-invasive noise probe has to minimize the disturbance \textcolor{dickred}{of the electronic energy distributions} via the heat leakage. In addition, its generated voltage noise, which contains the desired local information, has to dominate over the spurious voltage noise of the conductor itself. These criteria are naturally fulfilled provided the sensor resistance, $R$, is much higher than that of the conductor, $r$, which can be realized with numerous material combinations and, e.g., in scanning tunneling microscopy~\cite{Kemiktarak2007,Balatsky2012,Burtzlaff2015}.  Here, we probe the local noise of a micro-sized constriction in a macroscopic normal metal conductor with the help of a diffusive InAs nanowire (NW), which provides good ohmic contact and is characterized by a convenient resistance ratio of $R/r\sim10^3$. \textcolor{dickred}{The spatial resolution in this case is limited by the diameter of the NW.} We succeed to observe the shot noise behavior and the lack of electron-phonon relaxation within the constriction, an impossible task for sample-averaging noise approaches. Our approach extends local primary measurements towards a few 10\,K temperature range, which is not accessible for spectral sensitive approaches~\cite{Pothier1997,Giazotto_Pekola_2006,Altimiras2009,Hoffmann2009,Venkatachalam2012}.

The SEM image of the actual device used for local noise measurements is shown in Fig.~\ref{sketch}b. In the middle, an n-doped InAs NW ($\approx\,70\,\text{nm}$ in diameter and 2$\,\mu$m in length), grown by Au-assisted chemical beam epitaxy~\cite{Gomes2015} and drop casted on a $\rm SiO_2/Si$ substrate, is visible as a light-gray thin rod. The rest shows a Ti/Au metallic nanostructure around the NW, fabricated with the e-beam lithography. Two ohmic contacts between the NW and metal layers are  marked as \textit{test end} (red circle) and \textit{cold end} (blue circle). The ohmic contacts are realized by means of the two bar-shaped constrictions, each of them further connected to four metallic terminals. These terminals are designed to be broader and twice thicker than the constrictions, to minimize their  resistive heating during non-equilibrium driving and improve a thermal coupling to the substrate. In between, the device is additionally equipped with six plunger gates, which are kept grounded throughout the experiment. We control the non-equilibrium noise at the \textit{test end} position by varying the driving (heating) current $I_H$ which flows through the constriction H on the left hand side. $I_H$ can be switched between two current paths: from  the terminals 1 or 2 towards the grounded terminal, referred to as $\rm\perp$ in the following. \textcolor{dickred}{On the \textit{cold end}, all the terminals apart from N are kept floating throughout the experiment.}

\begin{figure}[t]
\begin{center}
\vspace{10mm}
  \includegraphics[width=0.8\columnwidth]{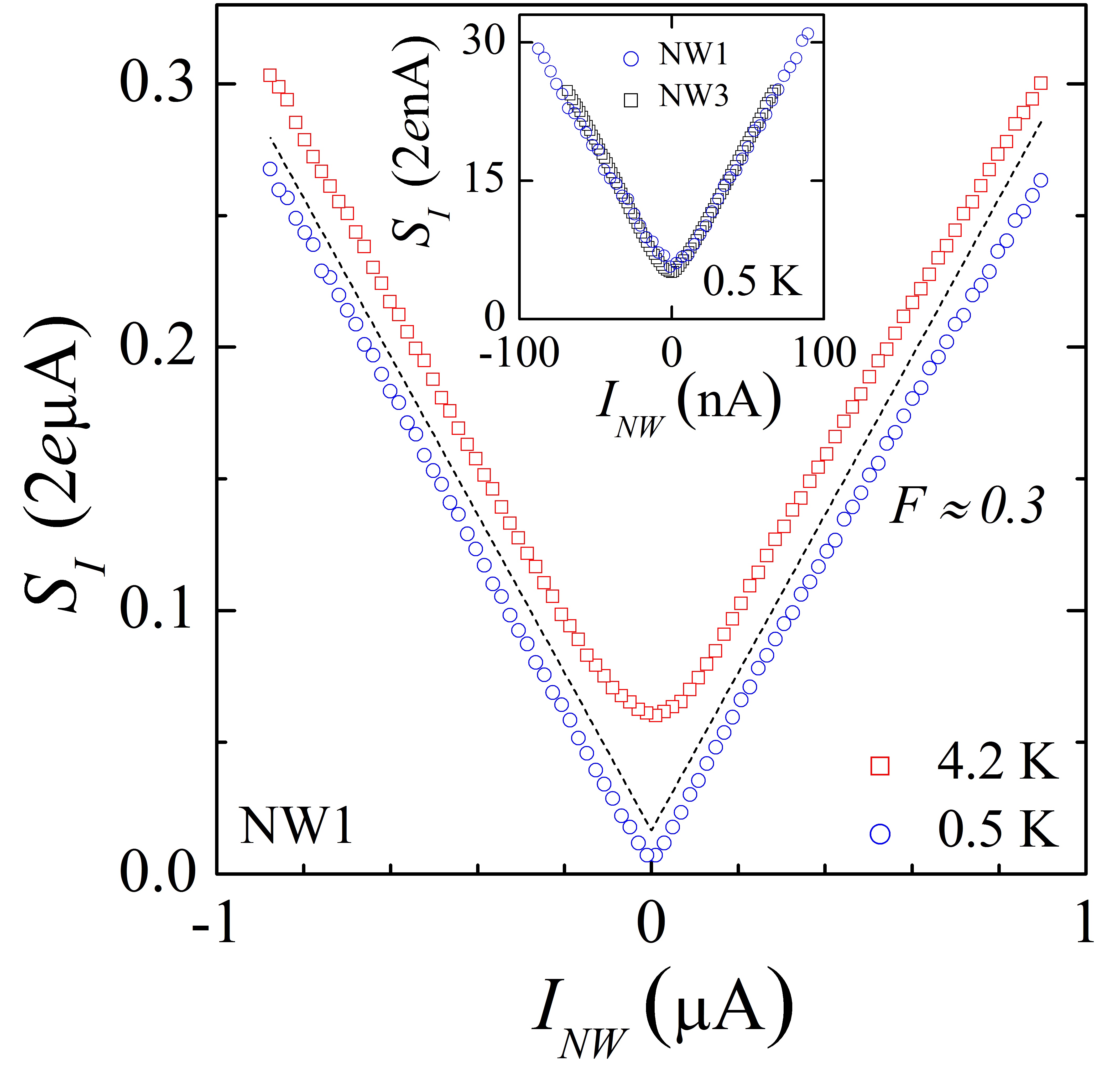}
   \end{center}
  \caption{Shot noise measurements. Noise spectral density in dependence on the bias current in device NW1. Different symbols correspond to different $T_0$, see legend. The slope of the dashed guide line corresponds to the Fano factor of $F=0.3$, close to the universal value for diffusive metallic conductors. Inset: magnified region of small $I_{NW}$ measured at $T_0\approx0.5\,$K  for the devices NW1 and NW3.}\label{shot_noise_all}
\end{figure}

\textbf{Shot noise calibration.}

The performance of the noise probe depends on the way the non-equilibrium fluctuations penetrate into the NW, as determined by the interplay of diffusion and inelastic scattering of the charge carriers~\cite{Blanter2000}. \textcolor{dickred}{Compared to the elastic transport regime and} depending on the nature of the inelasticity, the noise can be either enhanced, as for the electron-electron (e-e) collisions~\cite{Nagaev1995}, or suppressed, as for the electron-phonon (e-ph) relaxation~\cite{Nagaev1992}. As shown below, none of these effects is relevant to our NWs, which is an ideal case for the noise probe. To understand this better, we perform a primary calibration, biasing the NW with a transport current ($I_{NW}$) via terminal N and measuring the induced shot noise, see Fig.~\ref{sketch}b. In this experiment, the constrictions remain in thermal equilibrium at a given bath temperature $T_0$. 

The result of such calibration is shown in Fig.~\ref{shot_noise_all}, where we plot the dependence of the measured spectral density ($S_I$) of the current noise on $I_{NW}$ at different $T_0$. At increasing transport current, $S_I$ crosses over from the equilibrium Johnson-Nyquist value of $4k_BT_0/R$ to the closely linear dependence $S_I\approx2eF|I_{NW}|$, where $F$, $k_B$ and $e$ are, respectively, the Fano factor, the Boltzman constant and the elementary charge and $R$ is the linear response resistance of the NW. This observation evidences the shot noise behavior with the Fano factor of $F\approx0.3$ (dashed line), which is remarkably close to the universal value of 1/3 in metallic diffusive conductors with negligible inelastic scattering~\cite{Nagaev1992,Beenakker_Buettiker_1992}. These findings are independent of the device choice, see the inset of Fig.~\ref{shot_noise_all}, and $T_0$ and persist up to $|I_{NW}|\approx1\,\mu\text{A}$, i.e. the voltage drop of more than 10~mV across the NW. We conclude that the charge transport in our InAs NWs occurs via elastic diffusion, whereas the e-ph relaxation~\cite{Nagaev1992} and the e-e   collisions~\cite{Nagaev1995} can be safely neglected for carrier excess energies below 10~meV in respect to the Fermi level. Remarkably, this observation is even more favorable, than ab initio estimates of the inelastic scattering rates~\cite{ref2supplemental}.

The fact that the NW behaves as an elastic diffusive conductor facilitates the calibration of the probe sensitivity to the noise at its \textit{test end}. The randomness of the carrier diffusion process inside the NW gives rise to a complete loss of correlations between the average occupancy of the quantum state and the direction of the carrier momentum. Still, $\overline{n_k}$ retains the dependence on the carrier energy $\varepsilon$ characterized by the energy distribution function $\overline{n_k}\equiv f_\varepsilon$. For elastic diffusion, $f_\varepsilon$ obeys the Laplace's equation~\cite{Nagaev1992}, that is $\partial^2f_\varepsilon(x)/\partial x^2=0$, where $x$ is the coordinate along the NW. The solution of this equation is:
\begin{equation}
f_\varepsilon(x)=\left(1-\frac{x}{L}\right)f_\varepsilon(0)+\frac{x}{L}f_\varepsilon(L),\label{solution_f}\end{equation}
with the boundary conditions $f_\varepsilon(0)$ and $f_\varepsilon(L)$ set, respectively, by the energy distributions at the \textit{test end} and the \textit{cold end} of the NW. Spontaneous current fluctuations in a given cross-section of the NW can be conveniently described in terms of the local noise temperature, \textcolor{dickred}{which is defined as~\cite{Nagaev1992,Nagaev1995} } $T_N(x)=k_B^{-1}\int{f_\varepsilon(x)(1-f_\varepsilon(x))d\varepsilon}$. These fluctuations sum up~\cite{Nagaev1992} \textcolor{dickred}{such that the $T_N(x)$ averages along the device} to give response of the noise sensor: 
\begin{equation}
T_S=\int{T_N(x)\frac{dx}{L}}\,\text{,}\,S_I=\frac{4k_B}{R}T_S,
\label{spectral_density}
\end{equation}
where $T_S$ and $R$ are, respectively, the measured noise temperature of the NW-sensor and its resistance. 

In the regime of shot noise calibration, $f_\varepsilon(0)$ and $f_\varepsilon(L)$ represent the equilibrium Fermi distributions with the chemical potential difference determined by the voltage drop across the NW. Here, the eqs.~(\ref{solution_f}) and (\ref{spectral_density}) predict the well-known double-step distribution in the middle of the NW, see Fig.~\ref{sketch}c, and the universal $F=1/3$ for the shot noise Fano factor~\cite{Nagaev1992}. The regime of local noise sensing, addressed below, is different in two respects: (i) the unknown energy distribution at the \textit{test end} is non-equilibrium and  (ii) no net current flows through the NW, $I_{NW}=0$. Here, 
the eqs.~(\ref{solution_f}) and (\ref{spectral_density}) relate the measured noise of the sensor and the local noise temperature of the conductor under test. In the limit of strong non-equilibrium, $T_N(0)\gg T_0$, we arrive simply at $T_S\approx\alpha T_N(0)$, that is the  sensor directly measures the noise temperature of the conductor at the position of their mutual contact. As discussed above, elastic diffusive transport in our NWs persists up to at least 10~meV excitation energies, making our local noise thermometry well applicable within a few 10\,K temperature range.  A pre-factor $\alpha$ is slightly sensitive~\cite{Sukhorukov1999} to the shape of $f_\varepsilon(0)$, e.g., $\alpha=\frac{2}{3}$ for a double-step shaped distribution, similar to that sketched in Fig.~\ref{sketch}c, and $\alpha=\frac{1+\ln{2}}{3}\approx0.56$ for an equilibrium Fermi-Dirac distribution. For definiteness, the a-priori unknown energy distribution at the \textit{test end} is assumed below to be of equilibrium type, which is not important for any \textcolor{dickred}{of our} conclusions.

\textbf{Sensing the local noise.}

\begin{figure}[t]
\begin{center}
\vspace{10mm}
  \includegraphics[width=0.9\columnwidth]{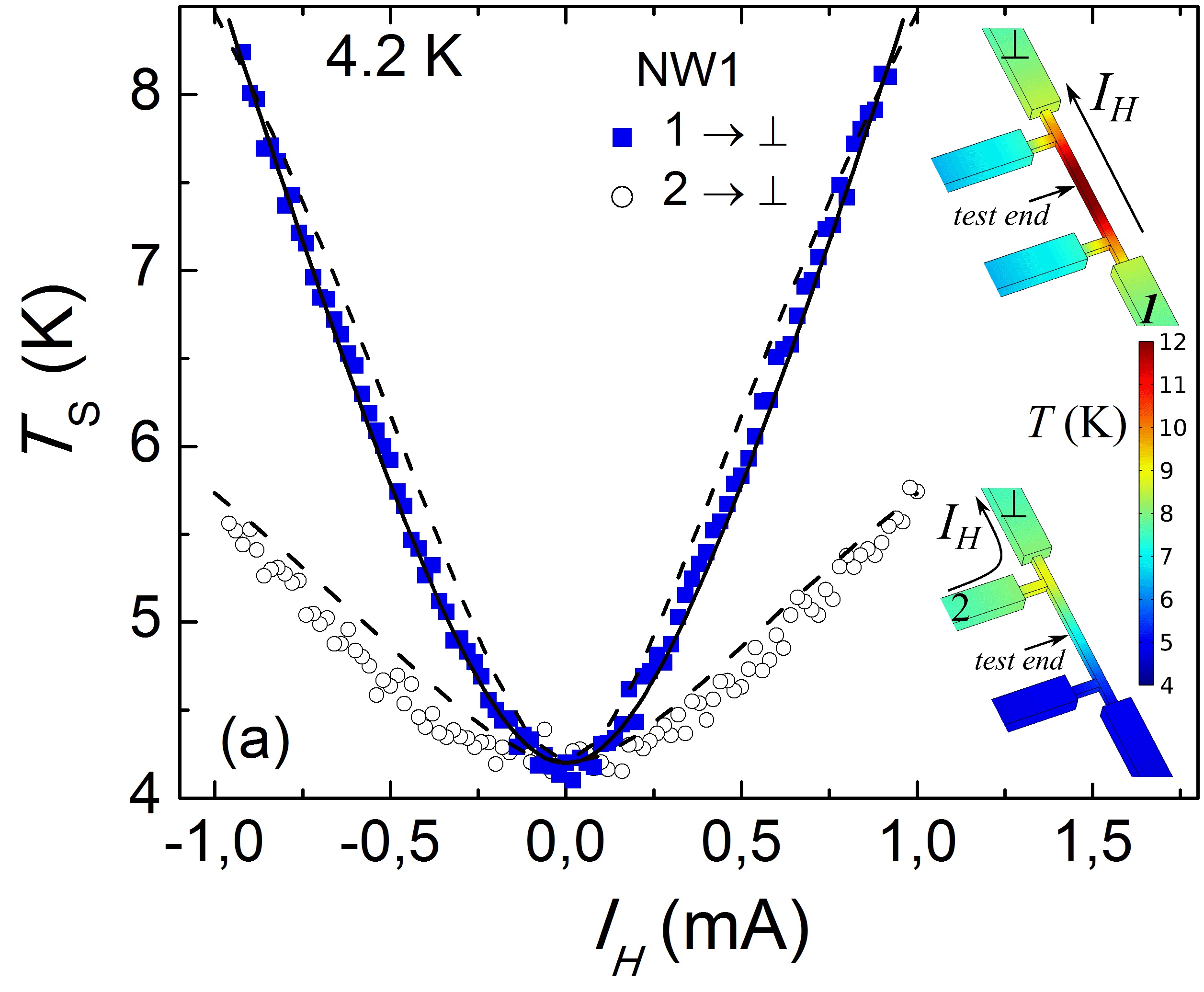}
   \end{center}
  \caption{Local noise measurement at $T_0=4.2$\,K. Measured noise temperature of the NW-sensor in dependence on the  current $I_H$ through metallic terminals for two biasing configurations shown by different symbols, see legend. Corresponding predictions of the model calculation are shown by the dashed lines. The shot noise fit in configuration $1\rightarrow\perp$ is shown by the solid line. Insets: Color-scale plots of the calculated spatial temperature distributions in the metallic constriction and neighboring terminals at $I_H=1\,$mA for the two configurations. The model accounts for the Joule heating, Wiedemann-Franz heat conduction and e-ph cooling and assumes local thermal equilibrium.}\label{R_heater4K}
\end{figure}
\begin{figure}[t]
\begin{center}
\vspace{10mm}
  \includegraphics[width=0.8\columnwidth]{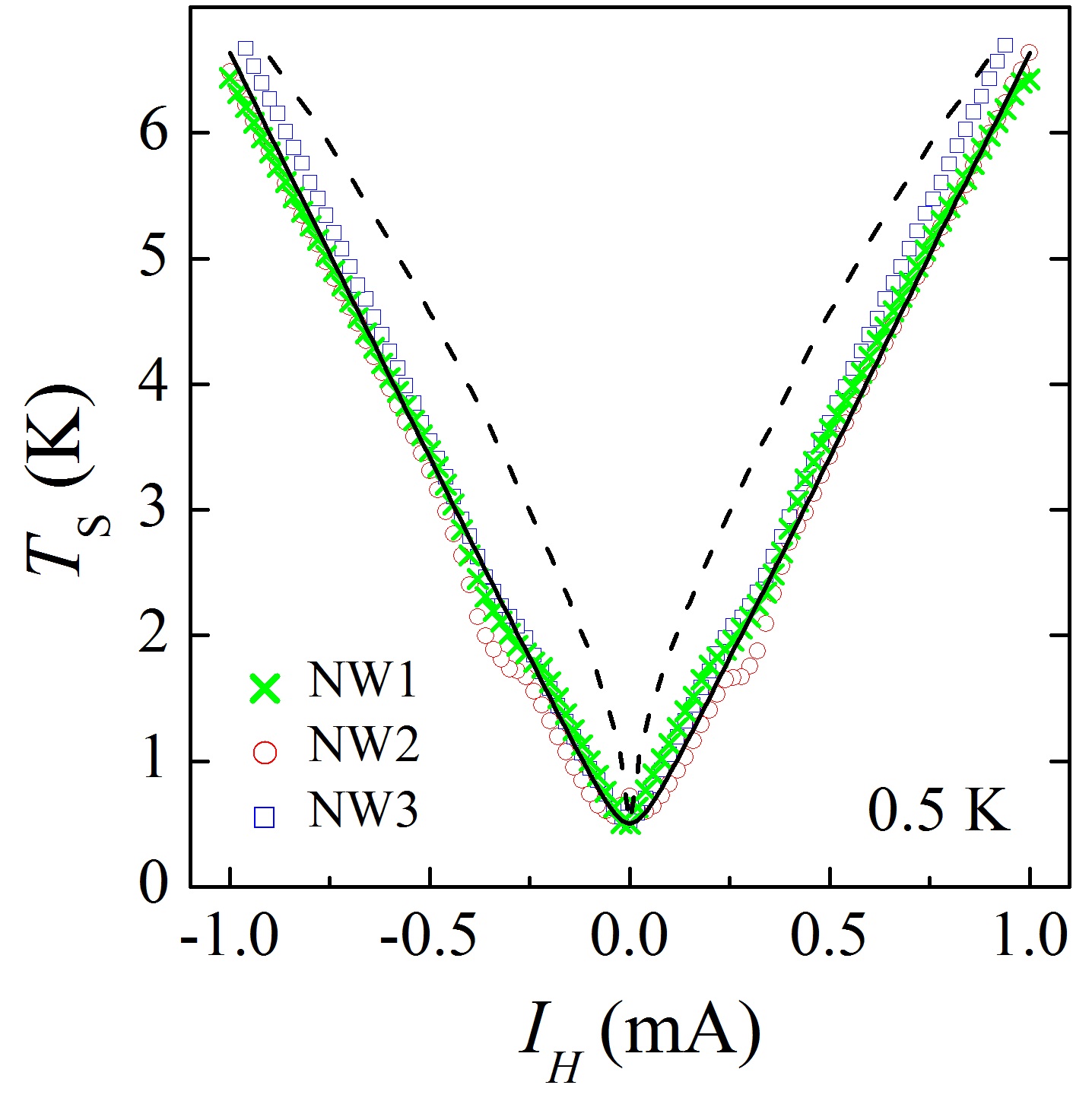}
   \end{center}
  \caption{Local noise measurement at $T_0=0.5$\,K. Measured noise temperature of the NW-sensor in dependence on the bias current $I_H$ in the configuration $1\rightarrow\perp$ for three devices used (symbols, see legend). Dashed line is the prediction of the numerical calculation at local equilibrium. Solid line corresponds to a shot noise fit.}\label{R_heater05K}
\end{figure}

Next we describe the main result of our paper --- the proof of concept of the local noise sensor. In this experiment, the noise of the unbiased NW-sensor is measured in response to the driving current $I_H$ sent via the constriction or bypassing it, respectively, in biasing configurations $1\rightarrow\perp$ and $2\rightarrow\perp$, see the micrograph in Fig.~\ref{sketch}b.

Fig.~\ref{R_heater4K} shows the measured noise temperature of the sensor $T_S$ in dependence on $I_H$ at 4.2~K for the device NW1 (solid  symbols) in the biasing configuration $1\rightarrow\perp$.  \textcolor{dickred}{By definition, the absolute values of $T_S$ translate into a measured voltage noise as $\overline{\delta V^2}\approx4k_BT_SR\Delta f$, where $\Delta f$ is the measurement bandwidth and $R\sim 10\,{\rm k}\Omega$ is the resistance of the NW, which is much higher than that of the metallic terminals connecting it to the ground, $r_M\sim 20\,\Omega$. On the other hand, $\overline{\delta V^2}$ consists of two contributions, the NW noise $\overline{\delta V_{NW}^2}$ and the terminals noise $\overline{\delta V_{M}^2}$. Obviously, $\overline{\delta V_{M}^2}\ll\overline{\delta V_{NW}^2}$, since the noise temperature of the metallic terminals, $T_M$, is the same order of magnitude as the temperature at the \textit{test end}, $T_N(0)$, and $T_N(0)\sim T_S$ as we derived above. Note also, that assuming $\overline{\delta V_{M}^2}\sim\overline{\delta V^2}$ one would arrive at an unreasonable estimate of $T_M\sim(R/r_M)T_S\sim1000\,$K. This remarkable feature of Fig.~\ref{R_heater4K} demonstrates that the voltage noise of the metallic terminals can be safely neglected and $T_S=T_{NW}$.} Therefore, in agreement with the eqs.~(\ref{solution_f}) and (\ref{spectral_density}), the measured noise is generated in the NW in response to a local non-equilibrium carrier distribution penetrating via its \textit{test end}. In comparison, the noise response is considerably weaker in a biasing configuration $2\rightarrow\perp$, for which  $I_H$ bypasses the constriction, shown by open circles in Fig.~\ref{R_heater4K}. This additionally verifies that the measured noise is a local characteristics.

The tendency towards a linear dependence of $T_S$ on the drive current for the upper trace in Fig.~\ref{R_heater4K} is consistent with
the shot noise \textcolor{dickred}{dominated} regime of the current biased constriction, \textcolor{dickred}{ which takes place at} inefficient electron-phonon (e-ph) energy relaxation. In this regime, under the assumption of local equilibrium, one expects a standard spatial temperature distribution along the constriction, determined by a balance of the Joule heating and the Wiedemann-Franz thermal conduction~\cite{Nagaev1995}. The maximum local temperature is attained at the {\it test end} and equals $T_N(0)=\sqrt{3e^2/(4\pi^2k_B^2)V^2+T_L^2}$, where $V=rI_H$ is the voltage drop on the constriction and $T_L$ is the temperature in the metallic leads. The solid line in Fig.~\ref{R_heater4K} corresponds to a fit parameter $r=3.7\,\rm\Omega$ and $T_L=T_0$, i.e., negligible Joule self-heating of the leads. Still, in this experiment the bath temperature is too high to rule out the impact of e-ph energy relaxation in the constriction. For comparison, we performed model calculations of spatial temperature distribution in the constriction and the leads for the two biasing configurations and with the Wiedemann-Franz heat conduction and the e-ph relaxation taken into account, see the insets of Fig.~\ref{R_heater4K}. Reasonable fits of the sensor response could be obtained for the e-ph cooling power density of $\Sigma(T^5-T_0^5)$, with $\rm\Sigma=0.1\,nW\mu m^{-3}K^{-5}$, see the dashed lines in Fig.~\ref{R_heater4K}.

The inefficiency of the e-ph relaxation becomes evident at a lower bath temperature. In Fig.~\ref{R_heater05K} we plot the results of the local noise sensing at $T_0=0.5\,$K in three devices, all in the biasing configuration $1\rightarrow\perp$. We observe a closely linear dependence on $I_H$ over an order of magnitude increase of $T_S$ (symbols). This behavior is qualitatively different from a sub-linear dependence of $T_S$ on $I_H$, which is expected in presence of e-ph relaxation, see the dashed line fit, obtained with the same parameter \textcolor{dickred}{$\rm\Sigma$} as the fits of Fig.~\ref{R_heater4K}. By contrast, the data is perfectly consistent with the linear shot noise behavior, see the solid line fit, which corresponds to the shot noise of the constriction with $r=3.6\,\rm\Omega$. Similar to the case of $T_0=4.2\,$K, this $r$ value is somewhat higher than typically measured four-terminal resistances of $2.8\div2.9\,\Omega$ in our devices, which is a reasonable discrepancy in view of the constriction geometry, see Fig.~\ref{sketch}b. The shot noise fit of Fig.~\ref{R_heater05K} fails to describe a little downward kink at $|I_H|\sim0.4\,$mA, more or less pronounced in all our devices. \textcolor{dickred}{The origin of this kink is not understood and might be related to peculiarities in the convection~\cite{Lee_Fairbank_1959} or even boiling of the surrounding liquid $^3$He.} The observation of shot noise behavior in Fig.~\ref{R_heater05K} demonstrates a success of our local noise measurement approach. Our short  constrictions are connected via macroscopic leads, each with a typical resistance of $\sim20\,\rm\Omega$. Thus, in conventional two-terminal noise measurements, even a minor increase of the Johnson-Nyquist noise of the leads, owing to their Joule self-heating, would completely mask the constriction shot noise~\cite{Henny1999}. 

\begin{figure}[t]
\begin{center}
\vspace{10mm}
  \includegraphics[width=0.8\columnwidth]{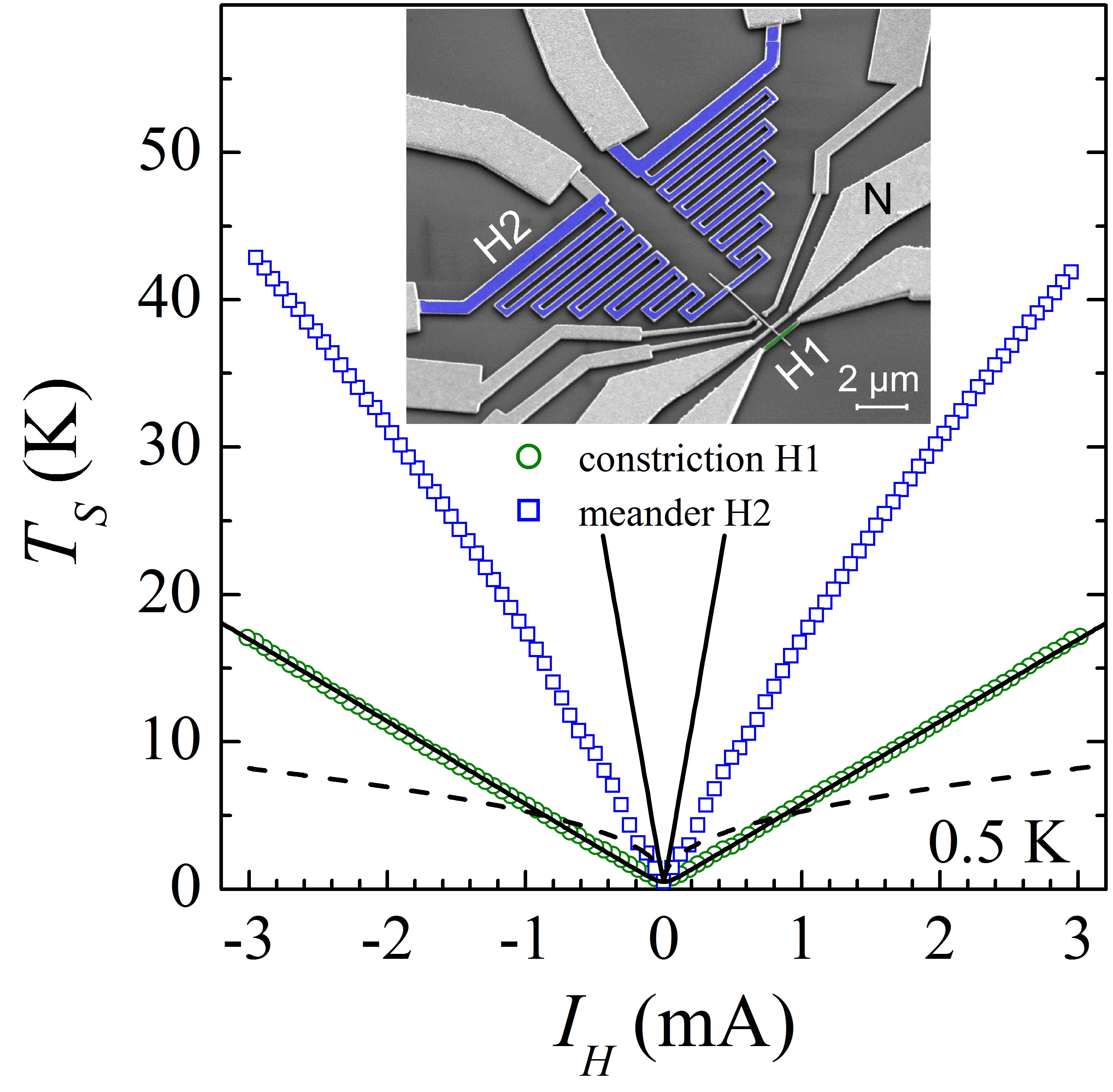}
   \end{center}
  \caption{Competition between the shot noise and the e-ph energy relaxation. Measured noise temperature of the NW-sensor in dependence on the bias current~$I_H$ through the short constriction~H1 (circles) and through the long meander-shaped heater~H2 (squares). Solid lines are the fits corresponding to the shot noise behaviour in H1 and H2 with $r=3.1\,\mathrm{\Omega}$ and $r=30\,\mathrm{\Omega}$ (four-terminal resistance of H2), respectively. Dashed line is the prediction of the e-ph energy relaxation case in~H2 with the cooling rate~$\Sigma=0.1\,\mathrm{nW\mu m^{-3}K^{-5}}$. Inset:  Colored SEM image of the device.
}\label{New_heater05K}
\end{figure}

The shot noise behavior in short constrictions is a consequence of hot electron out-diffusion in the leads, which occurs faster than the time-scale of e-ph relaxation, $\tau_{diff}\ll\tau_{e-ph}\sim T^{-3}$. An estimate based on a free-electron model~\cite{kaganov1957} is roughly consistent with the experiment, giving $\tau_{diff}\sim\tau_{e-ph}\sim150$\,ps at $T=10$\,K. In longer constrictions the diffusion time increases as $\tau_{diff}\sim L^2$, and we expect the e-ph relaxation to take over. This is demonstrated in a second generation devices, in which a NW is divided in two sections, each serving as a local noise probe, see the inset of Fig.~\ref{New_heater05K}. The local noise signals, which arise from biasing either a short constriction H1 or a long meander-shaped heater H2, are picked up at the contact N. These signals are plotted in the body of Fig.~\ref{New_heater05K}, symbols. The noise of the short constriction H1, cirlces, has a linear dependence on current, which is perfectly consistent with the shot noise fit for $r=3.1\,\rm\Omega$, see the lower solid line. Remarkably, the shot noise behavior is shown to persist here up to $I_H=3$\,mA, corresponding to local temperatures of $T_N(0)\approx30$\,K in the center of the constriction. By contrast, the noise of the long heater H2, squares, is much smaller than its expected shot noise value, upper solid line, and exhibits a sub-linear current dependence, evidencing a dominant role of the e-ph relaxation. Yet, in this case, the measured $T_S$ is considerably higher than the estimate \textcolor{dickred}{obtained under assumptions of the dominant e-ph relaxation, $P_J=\Sigma(T^5-T_0^5)$, where $P_J$ is the Joule power dissipated per unit volume of the heater H2}, see the dashed line in Fig.~\ref{New_heater05K}. \textcolor{dickred}{This indicates} that e-ph cooling is bottlenecked by Kapitza resistance and/or poor substrate heat conduction. From these data, we estimate that in our short constrictions the e-ph relaxation dissipates just a few percent of the total Joule heat, explaining the observed shot noise behavior.

In summary, we realized a local noise measurement in a current biased metallic conductor using InAs nanowire as a miniature noise probe. This approach enabled us to measure a shot noise of a micro-sized constriction connected to macroscopic resistive leads, an impossible task for conventional noise approaches. Our experiment represents a conceptually simple primary local measurement, not \textcolor{dickred}{limited} to very low temperatures \textcolor{dickred}{and small lead resistances} and compatible with various material combinations. As such, we envision, that local noise probing will turn useful for non-invasive spatially resolved studies in non-equilibrium nano-circuits.
 
We gratefully acknowledge discussions with S.G.~Kafanov, T.M.~Klapwijk and I.S.~Burmistrov. This work was supported in part by the
Russian Academy of Sciences, the CNR through the bilateral CNR-RFBR projects 2015–2017, the Ministry of Education and Science of the Russian Federation Grant No.14Y.26.31.0007, the RFBR Grants 15-02-04285 and 15-52-78023.

{\small \textbf{Methods.}  Au-assisted  Se doped InAs NWs are grown by chemical beam epitaxy on an InAs(111) B substrate. The carrier density of the InAs NWs derived by  field effect measurements is about $\rm 1\times10^{18} cm^{-3}$. \textcolor{dickred}{Typical ohmic contact resistance in our devices is below 100\,$\rm\Omega$, whereas the NW resistance is about 10\,k$\rm\Omega$ per micrometer.} We performed the measurements in two $^3$He inserts, with the samples immersed in liquid (at $T=0.5$\,K) or in gas (at $T=4.2$\,K) and placed vertically face down. The shot noise spectral density was measured using home-made low-temperature amplifiers (LTamp) with a voltage gain of about 10\,dB, input current noise of $\rm \sim10^{-27}\,A^2/Hz$ and dissipated power of $\sim200\,\mu$W. We used a resonant tank circuit at the input of the LTamp, see the sketch in Fig.~\ref{sketch}b, with a ground bypass capacitance of a coaxial cable and contact pads $\sim40\,$pF, a hand-wound inductance of $\sim6\,\mu$H and a load resistance of $\rm 10\,k\Omega$. The output of the LTamp was fed into the low noise 75\,dB total voltage gain room temperature amplification stage followed by a hand-made analogue filter and a power detector. The setup has a bandwidth of $\sim0.5$\,MHz around a  center frequency of $\approx10\,$MHz. A calibration was achieved by means of equilibrium Johnson-Nyquist noise thermometry. For this purpose we used a commercial pHEMT transistor connected in parallel with the device, that was depleted otherwise. All transport measurements were performed with the help of a two-terminal or four-terminal lock-in resistance measurement. }

{\small 
\textbf{Additional Information}.

\textit{Author Comtributions.} DE, FR and MR fabricated the devices. EST performed experiments and prepared the figures. EST, DVS and VSK analyzed the data. LS, SR and VSK have jointly supervised the work and written the manuscript. All authors reviewed the manuscript.

\textit{Competing financial interests.} The authors declare no competing financial interests.

}

\end{document}


\title{Local noise in a diffusive conductor. Supplemental Material.}
\author{E.S.~Tikhonov}
\affiliation{Moscow Institute of Physics and Technology, Dolgoprudny, 141700 Russian Federation}
\affiliation{Institute of Solid State Physics, Russian Academy of
Sciences, 142432 Chernogolovka, Russian Federation}
\author{D.V.~Shovkun}
\affiliation{Moscow Institute of Physics and Technology, Dolgoprudny, 141700 Russian Federation}
\affiliation{Institute of Solid State Physics, Russian Academy of
Sciences, 142432 Chernogolovka, Russian Federation}
\author{D.~Ercolani}
\affiliation{NEST, Istituto Nanoscienze -- CNR and Scuola Normale Superiore, Piazza S. Silvestro 12, I-56127 Pisa, Italy}
\author{F.~Rossella}
\affiliation{NEST, Istituto Nanoscienze -- CNR and Scuola Normale Superiore, Piazza S. Silvestro 12, I-56127 Pisa, Italy}
\author{M.~Rocci}
\affiliation{NEST, Istituto Nanoscienze -- CNR and Scuola Normale Superiore, Piazza S. Silvestro 12, I-56127 Pisa, Italy}
\author{L.~Sorba}
\affiliation{NEST, Istituto Nanoscienze -- CNR and Scuola Normale Superiore, Piazza S. Silvestro 12, I-56127 Pisa, Italy}
\author{S.~Roddaro}
\affiliation{NEST, Istituto Nanoscienze -- CNR and Scuola Normale Superiore, Piazza S. Silvestro 12, I-56127 Pisa, Italy}
\author{V.S.~Khrapai} 
\affiliation{Moscow Institute of Physics and Technology, Dolgoprudny, 141700 Russian Federation}
\affiliation{Institute of Solid State Physics, Russian Academy of
Sciences, 142432 Chernogolovka, Russian Federation}


\begin{abstract}
 
\end{abstract}

\maketitle

\section{Scattering times in nanowires}

The shot noise measurements of Fig.~2 of the main paper demonstrate a close to universal 1/3 value of the Fano factor, indicating that our InAs nanowires (NWs) are in the regime of elastic diffusive transport. This behavior persists for bias currents up to 1\,$\mu$A and 
bias voltages across the NW up to 10-20\,mV. Corresponding quasiparticle energies measured in respect to the Fermi energy
are about $\varepsilon\sim10$\,meV. In the following, we compare this experimental observation with the ab-initio estimates of the elastic vs inelastic scattering times. 

\textbf{Disorder scattering}

InAs NWs used in this work have a typical diameter of $d\approx70$\,nm (hexagonal cross-section, measured from corner to corner) and length between the contacts of $L\approx2\,\mu$m. The carrier density is about $n\approx10^{18}\,$cm$^{-3}$, that corresponds to a Fermi energy of $E_F\approx160\,$meV and the Fermi velocity at $v_F\sim1.6\times10^6\,$m/s. At liquid He temperatures the NW resistance is in the range $R\approx10-20\,{\rm k}\Omega$, hence a resistivity of ${\rm \rho\sim\pi d^2R/4L\approx 3\,m\Omega\cdot cm}$. This corresponds to a mean-free path for elastic disorder scattering $l_{mfp}=3\pi^2\rho^{-1}k_F^{-2}\hbar/e^2\sim 40\,\text{nm}$ and mobility of about $2\times10^3\,{\rm cm^2/Vs}$. We evaluate the quasiparticle dwell time in a NW at $\tau_{dwell}=L^2/v_Fl_{mfp}\sim60\,$ps.

\textbf{Electron-electron scattering}

We evaluate the electron-electron (e-e) scattering time based on the calculation of the quasiparticle life time in a 3D Fermi-liquid~\cite{Quinn1958}. With the Thomas-Fermi screening wave-vector defined as $q_{TF}=(4e^2k_F^2/\pi\hbar v_F)^{1/2}$ we obtain for the e-e scattering time: $\tau_{ee}= (\varepsilon/E_F)^{-2}(e^2/32/\hbar/q_{TF}k_F^2\pi)^{-1}\propto\varepsilon^2$. With $\varepsilon=2\,$meV this estimate results in $\tau_{ee}\sim 130\,$ps.

\textbf{Electron-phonon scattering} 

Following Ref.~\cite{Ridley1991} we express the energy relaxation rate per unit volume for deformation potential scattering on acoustic phonons in the Bloch-Gr$\rm\ddot{u}$neisen regime as $dE/dt=6D^2m^2(k_BT_e)^5(\pi^3\rho_m\hbar^7s^4)^{-1}$. Here $k_B$ is the Boltzman constant, $T_e$ is the electronic temperature, $D=4.9\,$eV, $s\approx4.3\,$km/s, $\rho_m\approx5.7\times10^3\,\text{kg/m}^3$ and $m\approx0.023\,m_e$ are, respectively, the deformation potential, typical sound velocity, mass density and the effective mass of the electrons in InAs. Here we assumed that the bath temperature is small compared to $T_e$.  The electron-phonon energy relaxation rate can be estimated 
as~\cite{Anderson1979} $\tau^{-1}_{e-ph}=(dE/dt)(C_eT_e)^{-1}$, where $C_e=(\pi/3)^{2/3}m\hbar^{-2}n^{1/3}(k_B)^{2}T_e$ is the electronic heat capacitance per unit volume. For a typical excess energy of $\varepsilon=k_BT_e$ we obtain $\tau^{-1}_{e-ph}=(dE/dt)(\pi/3)^{-2/3}\hbar^2m^{-1}n^{-1/3}\varepsilon^{-2}\propto\varepsilon^3$. With $\varepsilon=2\,$meV this results in $\tau_{e-ph}\sim 30\,$ps.

While the above estimates are consistent with the elastic diffusive transport in our InAs NWs for quasiparticle energies up to $\varepsilon=2$\,meV, one would expect the e-ph energy relaxation to come into play at higher excitations. This would manifest itself in a reduction of the shot noise spectral density from the universal Fano factor value $F=1/3$ already around $I\approx\varepsilon/eR\sim150\,$nA, which is not observed in our experiment, see Fig.2 of the main paper. Such an apparent weakening of the e-ph interaction in InAs NWs may result from screening of the e-ph interaction and/or reduced quasi-1D phase space for scattering. On the other hand, we can not exclude the weak e-ph energy relaxation originates from a bottle neck process of escape of the non-equilibrium phonons from the NW. In this case, the effective lattice temperature would follow that of the electronic system, and one could neglect the shot noise suppression. Still, the experimental $F\approx1/3$ would be a pure coincidence in such a situation.

\section{Scattering length scales in the contact metal}

Similarly, we estimated the length scales relevant for the metallic constriction, consisting predominantly of Au. Here we used $n\approx7\times10^{22}\,$cm$^{-3}$, $m\approx1.1\,m_e$, $E_F=5.6$\,eV,  $D=2E_F/3\,$, $s\approx3.2\,$km/s, ${\rm \rho\approx 3.2\,\mu\Omega\cdot cm}$ and $\rho_m\approx19.3\times10^3\,\text{kg/m}^3$. With these parameters, the dwell time in the constriction ($L\approx2\,\mu$m) is estimated at $\tau_{dwell}\sim130\,$ps, while the inelastic time scales for a quasiparticle energy of $\varepsilon=1\,$meV are $\tau_{ee}\sim 16\,$ns and $\tau_{e-ph}\sim 400\,$ps. This estimates correspond to a local thermalization length via e-e scattering of about $\sim25\,\mu$m and the energy relaxation length via e-ph interaction of about $\sim3.5\,\mu$m, both longer than $L$. Note, however, that while the inefficiency of the e-ph relaxation follows directly from the linear dependencies of $T_S$ on $I_H$ in Figs.~4 and~5 of the main paper, the role of the e-e processes remains hidden in our experiments. For example, even a tiny concentration of magnetic impurities might strongly accelerate thermalization via e-e processes, see e.g.~\cite{Glazman2001,Goeppert2001,Anthore2003}.

That $\tau_{dwell}<\tau_{ee}\, ,\tau_{e-ph}$ is consistent with the observation of linear dependence of the measured noise temperature on the driving current in Fig.~4 of the main article. Regarding e-ph interaction, the above parameters correspond to the e-ph energy relaxation rate of $\rm\Sigma\approx0.1\,nW\mu m^{-3}K^{-5}$, which is close to a free-electron model estimate~\cite{kaganov1957} and a factor of 5 smaller than the values used in the literature~\cite{Steinbach1996,Henny1999}. Possible discrepancy can be regarded minor in view of the strong energy dependence of $\tau_{e-ph}$.

\bibliography{InAsbbl}